\documentstyle[pra,aps]{revtex}
\def\up{\uparrow}
\def\dn{\downarrow}

\begin{document}
\preprint{\today}
\draft
%
%%%%%%%%%%%%%%%%%%%%%%%%%%%%%%%%%%%%%%%%%% TITLE PAGE
%
\title{Lindblad Approach to \\ 
Nonlinear Jaynes-Cummings Dynamics of a Trapped Ion} 
\author{Roberto Onofrio\thanks{E-mail: Onofrio@padova.infn.it} and 
Lorenza Viola\thanks{E-mail: Viola@mvxpd5.pd.infn.it} }
\address{Dipartimento di Fisica ``G. Galilei'', 
Universit\`a di Padova, and INFN, Sezione di Padova, \\ 
Via Marzolo 8, Padova, Italy 35131, \\
and INFM, Sezione di Roma 1, P. le A. Moro, 2, Roma, Italy 00185}
\date{\today}
\maketitle
%
%%%%%%%%%%%%%%%%%%%%%%%%%%%%%%%%%%%%%%%%%%%% ABSTRACT
%
\begin{abstract}
The Lindblad approach to open quantum systems is introduced for studying 
the dynamics of a single trapped ion prepared in nonclassical motional 
states and subjected to continuous measurement of its internal population. 
This results in an inhibition of the dynamics similar 
to the one occurring in the quantum Zeno effect. 
In particular, modifications to the Jaynes-Cummings collapses and revivals 
arising from an initial coherent state of motion in various regimes of 
interaction with the driving laser are \vspace{1cm}dealt in detail.
\end{abstract}
%
%%%%%%%%%%%%%%%%%%%%%%%%%%%%%%%%%%%%% PACS NUMBERS
%
\pacs{03.65.Bz, 42.50.-p, 32.80.Pj}  
%
%%%%%%%%%%%%%%%%%%%%%%%%%%%%%%%%%%%%% PAPER CONTENT
\begin{center}
(To be published in {\sl Physical Review A})
\end{center}

%\narrowtext
%\section{Introduction}
In recent years, several efforts have been made to clarify the 
role of the measurement process during the evolution of a quantum system. 
Besides the conceptual viewpoint, this also has a practical
relevance in predicting the results of experiments and possibly pointing out 
some working conditions where genuine quantum phenomena are expected 
\cite{GREEN}. Despite of the different languages, ranging from 
group algebraic and path integral techniques \cite{LINDBLAD,MENSKY} to 
stochastic evolution models \cite{DIOSI,BELAV}, which have been employed to 
analyze the quantum measurement problem, all these approaches 
recognize that a measured system is {\it not} closed but interacts with the 
environment schematizing the meter. An equivalence loop connecting these 
different descriptions has been established in \cite{MQM}, resulting in an 
unified picture which incorporates the measurement process through an 
unique parameter proportional the strenght of the system-meter interaction. 
In the present paper, we apply this scheme for investigating the influence of a 
continuous measurement process on the dynamics of a trapped and 
laser-irradiated ion. This system has lately attracted growing interest, 
culminating in the experimental generation of nonclassical motional states 
\cite{MEEKHOF,GATTI}. The possibility to read out the vibrational state of the 
ion following the evolution of its internal levels under a Jaynes-Cummings 
type interaction has been at the heart of the experimental procedure. As we 
shall see, by including the measurement process an inhibition of the two-level 
transitions dynamics emerges, which could be made observable via the 
quenching of collapses and revivals in a coherent-state prepared ion.

Our physical system is a single trapped ion interacting with a classical 
laser field and undergoing a continuous measurement of its internal 
population. In practice, one is dealing with a collection of independent 
evolutions of single ions starting from the same initial state and {\it 
averaged} results are actually registered as outcomes of the 
experiment. According to the general discussion of \cite{MQM}, we can either 
model each single history as a realization of a pure-state stochastic 
evolution, then averaging over the associated {\it selective} measurement 
results, or we can consider a deterministic ensemble evolution which directly 
leads to averaged {\it nonselective} predictions. Our choice of the latter 
method relies on the exact solvability of the model. The analysis is based on 
the master equation for the reduced density operator $\hat{\rho}(t)$ of the 
system, which results from tracing out the variables of the external 
environment from the overall system + environment density operator and 
is written in the Lindblad form 
\cite{LINDBLAD,MQM}: 
\begin{equation}
{d \hat{\rho}(t) \over dt}= - {i \over \hbar} 
[\hat{H}(t),\hat{\rho}(t)] - {\kappa \over 2} 
[\hat{A},[\hat{A},\hat{\rho}(t)]]\:,   
\label{MASTER}
\end{equation}
where $\hat{H}(t)=\hat{H}_0+\hat{H}_{int}(t)$ describes the dynamics of 
the unmeasured closed system and the Lindblad operator, representing in 
general the influence of the environment on the system, is related in this 
case to the measured observable $\hat{A}$. The parameter $\kappa$ expresses 
the strength of the coupling of the measured system to the meter \cite{MQM}, 
hereafter assumed to be time-independent. 
The Hamiltonian $\hat{H}_0$ describes the 1D-harmonic 
oscillator associated to the center-of-mass degree of freedom
and the two internal states of the ion:
\begin{equation}
\hat{H}_0= \hat{H}^{cm}+\hat{H}^{el}=\hbar \omega 
\, \hat{a}^{\dagger} \hat{a} + \hbar \omega_{21} \,\hat{\sigma}_z \:,
\label{HZERO}
\end{equation}
being $\omega$ the vibrational frequency, $\omega_{21}$ the two-level 
transition frequency, $\hat{a}$ and $\hat{\sigma}_z$ the boson annihilation 
operator and the pseudo-spin $\hat{z}$-operator respectively. The trap 
frequency is supposed large enough to neglect the atomic spontaneous emission 
(strong confinement limit \cite{MEEKHOF}). The interaction $\hat{H}_{int}(t)$ 
with the laser field is modeled by a nonlinear multiquantum Jaynes-Cummings 
model (JCM) which, in rotating-wave approximation, is written as 
\cite{VOGEL}
\begin{equation}
\hat{H}^{int}(t)={\hbar \Omega_0 \over 2} \,\mbox{e}^{i \omega_L t} \cos 
[\eta (\hat{a}+\hat{a}^{\dagger}) + \varphi] \, \hat{\sigma}_- + 
\mbox{H.c.}  \:, \label{HINT}
\end{equation}
where $\omega_L$ is the laser frequency, $\Omega_0$ the fundamental Rabi 
frequency, $\hat{\sigma}_-$ the lowering operator and 
$\eta= \omega_L \Delta x_{SQL} /c$ the Lamb-Dicke parameter related to 
the position standard quantum limit of the ion. In Eq. (\ref{HINT}) a 
standing-wave laser field is considered, the phase $\varphi$ fixing the 
position of the trap potential with respect to the wave. The 
measured observable $\hat{A}$ depends upon the specific experiment.
If the population of the internal ground state is measured, for instance by 
collecting the fluorescence emitted after stimulated transition to 
a third auxiliary level as done by Meekhof {\it et al.} \cite{MEEKHOF}, the 
operator $\hat{A}$ is the occupancy of level $\downarrow$, {\it i.e.}  
$\hat{A}=\hat{\sigma}_-\hat{\sigma}_+$ in our formalism. 

It is convenient to introduce the representation defined by the eigenkets 
of $\hat{H}_0$, $|S,n \rangle$, $S=\dn,\up,n=0,\ldots,\infty$, with 
matrix elements $\rho_{Sn,S'm}(t)=\langle S,n| \hat{\rho}(t) | S',m \rangle$.
Moreover, let us assume that the ion is in the low excitation regime 
($\omega \gg \Omega_0$) and the laser is tuned to the 
$k$-{th} vibrational sideband, {\it i.e.} $\omega_L=\omega_{21}+k \omega$, 
$k \in \mbox{Z}$, two conditions usually fulfilled in laboratory 
\cite{MEEKHOF}. This implies that transitions between states $|\dn,n \rangle$ 
and $|\up,n+k \rangle$, involving the exchange of $k$ vibrational quanta, 
are resonantly enhanced, whereas all the off-resonant couplings are rapidly 
oscillating with frequency $\omega$ and can be disregarded (JCM 
approximation \cite{VOGEL}). As a notable case, when 
$\varphi = \pm \, \pi/2$ and $\eta \ll 1$ (Lamb-Dicke limit), 
the well-known linear one-quantum JCM or anti-JCM operators 
$\eta(\hat{a}\hat{\sigma}_+ + \hat{a}^{\dagger}\hat{\sigma}_-$),  
$\eta(\hat{a}^{\dagger}\hat{\sigma}_+ + \hat{a}\hat{\sigma}_-$) 
are recovered by tuning to the first red or blue sideband, $k=-1$ or 
$k=1$ respectively  \cite{JAYNES}. 
In what follows we focus on the properties of the internal dynamics of the 
ion, which can be extracted from the reduced atomic density matrix,
$ \sigma_{SS'}(t)= \sum_{n=0}^{\infty} \rho_{Sn,S'n}(t)$.
We can therefore restrict ourselves to the subset of the density matrix 
$\rho_{Sn,S'm}$ having  $n=m$. According to the master equation (\ref{MASTER})
and for a generic $k$-quantum resonance, their time development is ruled by: 
\begin{eqnarray}
& & \left\{\begin{array}{l}
 \dot{\rho}_{\dn n \dn n}(t) = 
- i \Omega_{n n+k}/2 \left[ \rho_{\up n+k\dn n}(t) 
\mbox{e}^{i \omega_L t}- \rho_{\dn n\up n+k}(t) \mbox{e}^{-i \omega_L t} 
\right] \:, \\
\dot{\rho}_{\up n+k \up n+k}(t) = 
- i \Omega_{n n+k}/2 \left[\rho_{\dn n\up n+k}(t) 
   \mbox{e}^{-i \omega_L t}  - \rho_{\up n+k\dn n}(t) \mbox{e}^{i \omega_L t}
   \right] \:,  \\
 \dot{\rho}_{\dn n \up n+k}(t) =  
\left(i \omega_{21} - \kappa/2 \right) \rho_{\dn n\up n+k}(t) 
- i \Omega_{n n+k}/2 \left[ \rho_{\up n+k\up n+k}(t)
 - \rho_{\dn n\dn n}(t) \right] \mbox{e}^{i \omega_L t} \:,   \\
 \dot{\rho}_{\up n+k \dn n}(t)  =  \left( -i\omega_{21} - \kappa/2 \right) 
\rho_{\up n+k\dn n}(t)  - i \Omega_{n n+k}/2 \left[\rho_{\dn n\dn n}(t) 
     - \rho_{\up n+k\up n+k}(t) \right]\mbox{e}^{-i \omega_L t} \: , 
\end{array}  \right. 
\label{EQS}
\end{eqnarray}
where, as usual, the nonlinear $k$-quantum Rabi frequencies 
expressed in terms of the Laguerre polinomials ${\cal L}^k_n$
\begin{equation}
\Omega_{n n+k}=
\Omega_0  \langle n | \cos[ \eta (\hat{a}+\hat{a}^{\dagger})+
\varphi]|n+k \rangle= \Omega_0 \left[\mbox{e}^{i\varphi} + 
(-1)^k \mbox{e}^{-i \varphi}\right](i\eta)^k 
{\left({n! \over {(n+k)!}} \right)}^{1/2} \mbox{e}^{-\eta^2 /2} 
{\cal L}^k_n(\eta^2)
\end{equation}
have been introduced \cite{VOGEL}. In the Lamb-Dicke regime 
($\eta \ll 1$) they reduce to the linear ones, {\it e.g.} 
$\Omega_{n n+1}= \eta \Omega_0 \sqrt{n+1}$ for one-quantum interaction.  
The following solution of (\ref{EQS}) holds for $n \ge -k$:
\begin{eqnarray}
\rho_{\dn n \dn n}(t) & = & 
  {1 \over 2} \Big( \rho_{\up n+k \up n+k}+\rho_{\dn n \dn n} \Big)(0) 
 -  {1 \over 2} \mbox{e}^{- \kappa t/4} 
\bigg\{ 
\Big( \rho_{\up n+k \up n+k}-\rho_{\dn n \dn n} \Big)(0)
\Big[ \cos(w_{n n+k}t) \nonumber \\
& + & {\kappa \over 4 w_{n n+k}} \sin(w_{n n+k}t) \Big]
+ i \, \mbox{Im}( \rho_{\dn n \up n+k}(0) ) 
{2\Omega_{n n+k} \over w_{n n+k}} \sin(w_{n n+k}t) \bigg\} \:, 
\nonumber \\  
%%%%% end 1
\rho_{\up n+k \up n+k}(t) & = & 
\Big( \rho_{\up n+k \up n+k}+\rho_{\dn n \dn n} \Big)(0) - 
\rho_{\dn n \dn n}(t)  \:,  \label{SOLUTION} \\
%%%%% end 2
\rho_{\dn n \up n+k}(t) & = & 
\mbox{e}^{i k \omega t -  \kappa t/4} 
\bigg\{ 
\mbox{Re}(\rho_{\dn n \up n+k}(0))\mbox{e}^{- \kappa t/4}
 +  \mbox{Im}(\rho_{\dn n \up n+k}(0))
\Big[ \cos(w_{n n+k}t) \nonumber \\
& - & {\kappa \over 4 w_{n n+k}} \sin(w_{n n+k}t) \Big]
- i \Big( \rho_{\up n+k \up n+k}-\rho_{\dn n \dn n} \Big)(0)
{\Omega_{n n+k} \over 2 w_{n n+k}} \sin(w_{n n+k}t) 
\bigg\}  \:,  \nonumber 
\end{eqnarray}
where  
\begin{equation}
w_{n n+k}=\sqrt{{\Omega^2_{n n+k}}-{\kappa^2 \over 16}} \:
\label{SYMB}
\end{equation}
and ${\rho}_{\up n+k \dn n}(t)={\rho}^*_{\dn n \up n+k}(t)$. Eqs. 
(\ref{SOLUTION}), holding within every {\it fixed} Jaynes-Cummings manifold 
$|\dn,n\rangle$, $|\up, n+k\rangle$, are formally equivalent to Eqs. 
(84)-(85) of Ref. \cite{MQM}, describing a two-level system subjected to 
a continuous occupancy measurement. When $\kappa=0$ the effect of the 
measurement disappears and oscillations with angular frequency 
$\Omega_{n n+k}$ result for fixed $n$. In the opposite limit of strong 
measurement coupling the frequency (\ref{SYMB}) becomes imaginary and 
an overdamped regime occurs in which transitions are inhibited, the 
so-called quantum Zeno effect. The borderline between the two regimes is 
ruled by the threshold $w_{n n+k}=0$, corresponding to a critical 
measurement coupling $\kappa^{crit}_{n n+k}=4 \Omega_{n n+k}$. In a general 
situation where different vibrational subspaces are involved, the 
$n$-dependence of this critical value globally results in a smoother 
transition.
The measured occupancy $P_{\dn}(t)=\sigma_{\dn\dn}(t)$ 
is straightforwardly evaluated from Eqs. (\ref{SOLUTION}) once the 
initial conditions are specified. We assume here, according to \cite{MEEKHOF}, 
that no entanglement between the internal and motional degrees of freedom 
occurs and only the ground internal level is populated, yielding
\begin{equation}
P_{\dn}(t)={1 \over 2} \bigg\{ 1+ \mbox{e}^{-\kappa t/4}
\sum_{n=0}^{\infty} \rho^{cm}_{nn}(0) \bigg[ 
\cos(w_{n n+k}t)+{\kappa \over 4 w_{n n+k}} \sin(w_{n n+k}t) 
\bigg] \bigg\}  \:,      \label{PDOWN}
\end{equation}
the matrix elements $\rho^{cm}_{nn}$ characterizing the number state 
distribution of the vibrational motion. The presence of entanglement, 
crucial to extend the description to Schr\"{o}dinger cat states \cite{GATTI}, 
can be included as well. Eqs. (\ref{PDOWN}) displays the effect 
of the measurement for an arbitrary vibronic distribution, namely an overall 
damping of the signal amplitude and, for each vibrational component, both a 
frequency shift (\ref{SYMB}) and a $\kappa$-weighted term. 

Let us now specialize the internal state evolution (\ref{PDOWN}) to an 
initial coherent distribution, 
$\rho^{cm}_{nn}(0)=\overline{n}^n\mbox{e}^{-\overline{n}}/n !$.   
In the limit in which no measurement is present ($\kappa=0$) and 
a linear Jaynes-Cummings interaction is realized by a sufficiently small 
Lamb-Dicke parameter, $P_{\dn}(t)$ undergoes collapses and revivals, a 
well-known quantum phenomenon due to dephasing and rephasing between the 
various Rabi oscillations, as firstly predicted by Eberly {\it et al.} 
\cite{EBERLY} and recently observed in cavity QED by using Rydberg atoms 
\cite{HAROCHE}. 
Fig. 1a shows the collapses and revivals occurring in a quantized trap, 
the center-of-mass being in a coherent state of average quantum number 
$\overline{n}$. The modifications intervening when the same system is coupled 
to the meter are depicted in Fig. 1b for a strength 
$\kappa =10^{-1} \kappa^{crit}_{01}$, in units of a reference 
value $\kappa^{crit}_{01}=4\Omega_{01}$. Provided 
$\kappa \ll \kappa^{crit}_{\tilde{n} \tilde{n}+k}$, 
$\tilde{n}=\mbox{Int}(\overline{n})$, an approximate evaluation of the 
series in (\ref{PDOWN}) leads to 
\begin{equation}
P_{\dn}(t) \simeq  1+ \cos(\eta \Omega_0 \sqrt{\bar{n}+1} \,t) 
\exp\bigg( {- {\kappa \over 4}t- {\eta^2 \Omega_0^2 \over 8} 
{\bar{n} \over {\bar{n}+1}} t^2} \bigg)
\:,      \label{PAX}
\end{equation}
which holds for a timescale $t \ll 2 \sqrt{\bar{n}}/\eta \Omega_0$, still 
including all the collapse evolution, and generalizes the result already 
quoted in \cite{EBERLY} for $\kappa=0$. Thus, even in this weak coupling 
regime, the measurement has a leading influence at short times due to 
its linear dependence in the exponent of the envelope in (\ref{PAX}). 
Similar considerations can be repeated for the nonlinear multiquantum
Jaynes-Cummings interaction, which is restored outside the Lamb-Dicke limit 
and has been extensively studied by Vogel and de Matos-Filho in case 
$\kappa=0$ \cite{VOGEL}. Two representative examples, corresponding 
to one-quantum and two-quantum couplings are shown in Figs. 2 and 3 
respectively. The parameters have been chosen to reproduce, in the 
closed-system limit, the experimental results reported by \cite{MEEKHOF} 
for a nonlinear one-quantum interaction and the same reference ratio 
$\kappa/\kappa^{crit}_{0k}=10^{-1}$, $k=1,2$, has been maintained as before. 
In all the cases the measurement tends to wash out collapses and revivals 
making more difficult to infer the quantized nature of the vibrational 
motion in the trap. For larger values of $\kappa/\kappa^{crit}_{0k}$ the 
suppression is even more pronounced, while for $\kappa/\kappa^{crit}_{0k} > 1$ 
a complete freezing of the population dynamics to the initial occupation  
is achieved, regardless of the initial vibrational state.
A qualitatively similar quenching of the collapse-revival phenomena for the 
linear one-quantum Jaynes-Cummings dynamics of a single two level atom in 
an optical cavity is described in \cite{TRAN} by including both spontaneous 
emission and cavity damping mechanisms. The relative insensitivity to 
spontaneous emission damping reported there further enforces the 
possibility of neglecting such effect in our description. 

Trapped ions seem particularly adequate to investigate the vanishing of 
revivals in coherent states induced by the measurement of their occupancy 
probability. Relaxation mechanisms due to spontaneous 
decay of the electronic transition or mechanical dissipation of the 
ion vibrational motion are indeed negligible \cite{ITANO,BROWN}.  
A technical source of decoherence has been instead identified in 
\cite{MEEKHOF} and ascribed to laser and frequency trap fluctuations, 
contributing to a finite linewidth for the electronic transition 
depending upon the vibrational state, $\gamma_n=\gamma_0 (n+1)^{0.7}$,
$\gamma_0=11.94$ kHz. This yields an average decay time 
$\tau_{\overline{n}}=\gamma_{\overline{n}}^{-1}=31.2\,\mu$s for 
$\overline{n}=3.1$. On the other hand the data collected in the quantum Zeno 
experiment \cite{ITANO}, exploiting the same auxiliary transition
$\mbox{}^2S_{1/2}\rightarrow \mbox{}^2P_{3/2}$ of $\mbox{}^9\mbox{Be}^+$ 
also used in \cite{MEEKHOF}, are fitted through  Eq. (\ref{MASTER}) 
with $\kappa \approx 4.9\cdot 10^4\,\mbox{s}^{-1}$ \cite{MQM}. 
The corresponding decay time is $\tau=4 \kappa^{-1} \simeq 816\,\mu$s,  
one order of magnitude larger than the technical decoherence decay constant, 
suggesting that an improvement in the stability performances of the 
experimental setup by roughly a factor $10^2$ should imply a 
measurement-dominated decoherence. The previous numerical 
example corresponds to a ratio 
$\kappa/\kappa^{crit}_{\tilde{n} \tilde{n}+k}=1.1\cdot 10^{-2}$, 
{\sl i.e.} a weak coupling regime. An increase of this ratio by two
orders of magnitude, for instance attainable by using the same measurement 
configuration but a smaller Rabi frequency $\Omega_0$, allows one to test the 
region where neither a closed system evolution nor a von Neumann instantaneous 
projection of the state are suitable for reproducing the experimental results. 
A description in terms of the Lindblad equation (\ref{MASTER}) is instead 
mandatory. 

The predicted effect does not exhaust the influence of the measurement 
process on the ion dynamics. As investigated in \cite{ONVIO} 
in a different context, the measurement also {\it indirectly} affects the 
average 
vibrational motion of the trapped ion, a quantum damping without classical 
analogue whose features will be described in \cite{JCMVO}.
Furthermore, analogous decoherence effects can be studied within the 
same framework in other physical systems, atomic Bose-Einstein condensates 
being particularly appealing in connection with a recent analysis of their 
collapses and revivals dynamics \cite{WRIGHT}. From a general viewpoint, 
experimental investigations of the effect of the measurement process as the 
one proposed here will allow for a quantitative description of this ultimate 
source of decoherence, a topic which besides its established importance in 
defining the boundary between quantum and classical worlds \cite{ZUREK} is 
becoming crucial for the issue of quantum computation \cite{CIRAC1}.

%%%%%%%%%%%%%%%%%%%%%%%%%%%%%%%%% ACKNOWLEDGMENTS

%\acknowledgments 
%This work has been supported by INFN, Sezione di Padova, and INFM, 
%Sezione di Roma 1, Italy.

%%%%%%%%%%%%%%%%%%%%%%%%%%%%%%%%% REFERENCES LIST

%
%%%%%%%%%%%%%%%%%%%%%%%%%%%%%%%%% FIGURE CAPTIONS
%
\begin{figure}

\caption{\label{fig1}
(a) Ground state occupation versus time for a linear one-quantum
resonance and an initial coherent state distribution, showing collapses and 
revivals. The parameters $\bar{n}=3.1$, $\eta=10^{-2}$ and 
$\Omega_{01}/2\pi= 4.75$ KHz have been used and $\kappa=0$. (b) Same as in 
(a) except $\kappa/\kappa^{crit}_{01}=10^{-1}$.}
\end{figure}
  
\begin{figure}
\caption{\label{fig2}
(a) The same as in Fig. 1a but for an one-quantum nonlinear Jaynes-Cummings 
model, $\Omega_{01}/2 \pi=94$ KHz and $\eta=0.202$. 
(b) $\kappa/\kappa^{crit}_{01}=10^{-1}$.} 
\end{figure}

\begin{figure}
\caption{\label{fig3}
(a) The same as in Fig. 2a but for a two-quantum nonlinear interaction, 
$\Omega_{02}/2 \pi=13.4$ KHz. (b) $\kappa/\kappa^{crit}_{02}=10^{-1}$.}

\end{figure}
\end{document}